\documentstyle[twocolumn,aps]{revtex}

\begin{document}

\hyphenation{Ka-pi-tul-nik}

\twocolumn[
\hsize\textwidth\columnwidth\hsize\csname@twocolumnfalse\endcsname
\draft

\title{Nonlinear I-V, High Flux-Flow Velocity Instability and Evidence for
Quantum Fluctuations in BSCCO Superconducting Films}
\author{J. Chiaverini$^1$, J.N. Eckstein$^2$, I. Bozovic$^3$, S. Doniach$^1$, and A. Kapitulnik$^1$ }
\address{$^1$Departments of Applied Physics and of Physics, Stanford University, Stanford, CA 94305, USA\\
$^2$Department of Physics, University of Illinois at Urbana Champaign, IL 61801, USA\\
$^3$Oxxel GMBH, D-28359 Bremen, Germany }

\date{\today}
\maketitle

\begin{abstract}

The large flux-flow velocity behavior of the high temperature
superconductor BSCCO in the mixed state is studied by measurement
of the current-voltage characteristic as a function of temperature and
magnetic field direction.  We find strong evidence for vortex-antivortex
pairs that are excited by quantum fluctuations and
contribute to the dissipation when dissociated by a large current.
\end{abstract}

]

Type II superconductors in the mixed state exhibit non-linear
current-voltage dependence associated with the long
energy-relaxation time of the conduction electrons \cite{larkin1}.
At high flux-flow velocities this non-linear behavior gives rise
to an instability which manifests itself as a distinct break in
the current-voltage characteristic. This behavior originates from
the nonequilibrium distribution of quasiparticles in a
superconductor driven by an electric field \cite{eliashberg} and
was first predicted by Larkin and Ovchinikov (LO)
\cite{larkin1,larkin2}. In the LO theory, as the vortex velocity
increases, quasiparticles exit the core \cite{kramer} and the
viscous drag is reduced, driving more quasiparticles from the
shrinking core.  When all quasiparticles leave the vortex core the
resistance of the sample jumps to a value close to the normal
state resistance. The voltage, $V^*$, at which this instability
occurs defines a critical velocity for the vortices, ${\bf
v}^*_\varphi$, independent of the magnetic field, ${\bf H}$, via
the relation:

\begin{equation}
\label{vstar} V^*=-({\bf v}^*_\varphi \times {\bf H} )\cdot {\bf L}
\end{equation}

\noindent where ${\bf L}$ is a vector along the direction of the
current flow whose magnitude is equal to the length of the current
path. This phenomenon has been observed in both low-$T_c$
\cite{musienko,samoilov,doettinger1} and high-$T_c$
\cite{doettinger1,doettinger2,doettinger3} superconductors. Since
the main assumption in the LO theory is that the distribution of
quasiparticles is spatially uniform, the theory may be valid only
near $T_c$ or at relatively high fields where the field is
homogeneous in the sample. This requires that the inelastic
scattering length of the quasiparticles at the critical velocity
be larger than the intervortex distance, $\l{_H}$
\cite{doettinger4}: $(D \tau_{in})^{1/2} \gg \l{_H}\cong (\Phi
_0/H)^{1/2}$; here $D$ is the quasiparticle diffusion constant,
$\tau_{in}$ is the inelastic scattering time, and $\Phi_0$ is the
flux quantum. Indeed, Doettinger {\it et al.} \cite{doettinger4}
estimated that for the case $(D \tau_{in})^{1/2} \ll \l{_H}\cong
(\Phi _0/H)^{1/2}$, the result for the critical velocity will be
modified, yielding a magnetic field dependence for which $
v^*_\varphi \equiv \mid {\bf v}^*_\varphi\mid \sim {(\Phi _0 /
H)^{1/2}/ \tau_{in}  }$. While this formula seems to fit the data
in the range of fields studied in \cite{doettinger4}, new lower
field data shows strong deviations.

In the present paper we investigate further the $I-V$
characteristics of BSCCO single crystal films in the temperature
range $T_c$/2. By using both field dependence and angular
dependence at a fixed magnetic field we confirm the vector product
nature of equation \ref{vstar}. Measuring at very low perpendicular fields,
we show that $V^*$ across the sample is
finite and independent of magnetic field down to $H=0$ with very
weak temperature dependence. This observation indicates the
existence of a large number of quantum mechanically excited
vortex-antivortex pairs, detected in our experiment by current
induced dissociation of the pairs followed by flux flow.

The samples were $300$ {\AA} thick BSCCO films grown on
$\langle100\rangle$ LaAlO$_3$ using layer-by-layer molecular beam
epitaxy, with resistively measured $T_c$'s of 82 K and a
resistance linear in $T$ above $T_c$. The films were c-axis
oriented, hence $I-V$ measurements represent properties of the a-b
plane. Sample composition was carefully controlled to ensure the
absence of second-phase defects \cite{eckstein}. High-resolution
photoemission on films grown under the same conditions show normal
and superconducting spectra consistent with the bulk-crystal
analog \cite{marshall}.

A split-coil magnet was used to allow the variation of the angle
between the plane of the sample and the field with a resolution of
0.05 degrees. Experimental procedure consisted of single
triangular wave sweeps of the current and simultaneous measurement
of the voltage across the sample.  Sweep rates were varied over
$\sim$1.5 orders of magnitude with no apparent effect to the
samples' behavior; this is evidence that sample heating did not
occur at these sweep rates. Also, the self-field at the
highest currents used is too small to explain any of the
results discussed below. Figure 1 depicts a typical sweep at
$T=35$ K where the instability is clearly visible and strong
hysteresis is observed. In what follows the values of $V^*$ and
$I^*$, the instantaneous current at which the instability occurs,
were measured on the rising side of the current pulse. Also
apparent in figure 1 is the strong non-linear regime before the
instability.  This nonlinear behavior will be discussed later.

As noted above, the instability voltage was measured for several
values of the magnetic field varying the angle $\theta$ between
the plane of the sample and the magnetic field from 0 to 90
degrees at various temperatures.  Figure 2 depicts a set of curves
taken at three different field magnitudes as a function of the
angle. Multiplying the applied field by $sin\theta$, we plot in
the inset of figure 2 the instability voltage as a function of
this perpendicular component only. The collapse of all points onto
a single curve in the whole range of fields up to 2.0 T is direct
evidence for the validity of the triple-product in equation
\ref{vstar}.  It also suggests that $V^*$ is controlled by vortex
dissipation. Through variation of the angle of the field, very
small values of the perpendicular component of the field can be
applied to the sample. As the point where $V^*$ is minimum is
approached, a remarkable result is obtained:  $V^*$ is not zero at
$H_\perp \equiv Hsin\theta$=0. In fact, careful examination of the
results presented in ref. \cite{doettinger4} suggests that this
phenomenon is also observed in YBCO thin films at sufficiently low
temperatures (this is manifested by the fact that plotting
$v^*_\varphi$ as a function of $H^{-1/2}$ shows deviations towards
saturation at low fields).

The observation of a finite $V^*$ at low fields suggests that the
extraction of a critical velocity from the $V^*$ data by using
equation \ref{vstar} ({\it i.e.} dividing the $V^*$ data by the
perpendicular component of the field) may obscure the important
features at low fields. However, it is evident from figure 2 that
past a well-defined low field, $V^*$ indeed increases sub-linearly
as a function of the field's perpendicular component similar to
what is suggested by \cite{doettinger4}. We find that the most
striking way to present the data is by plotting $V^*$ as a
function of the square root of the perpendicular component. This
is done in figure 3 where we present data for three temperatures
in the range of $\sim T_c/2$. The data clearly show a linear
dependence of $V^*$ in $H_\perp^{1/2}$ at higher fields that
saturates to a finite $V^*$ value at low fields. It is interesting
to note that the $H_ \perp ^{1/2}$ dependence extrapolates to
$V^*$=0 from this higher field regime. In other words, at higher
fields the system does not know that it will have a finite value
of $V^*$ at low fields. Also, in no region of fields was there
found a crossover to $V^*$ linear with field as predicted by the
LO theory \cite{larkin1}. While for lower temperatures the
saturation value decreases,  it is evident from the data that this
saturation remains finite even at $T=0$ K.

The above results suggest that vortex dissipation is observed at
low temperatures and zero magnetic field. Such an effect can be
due to vortex-antivortex pairs that are excited by some mechanism.
Because the thermal energy is low at $\sim T_{KT}/2$, thermally
excited vortex pairs are excluded  \cite{corson1}. At 25 K {\it
e.g.} we expect a density less than $10^{-7}$ of thermally excited
vortices. Moreover, the sample is very clean and pairs created by
static disorder can be ruled out. This can also be determined from
the data obtained at higher fields: $V^*$ extrapolates to zero (no
dissipation) as $H_{\perp}^{1/2}$ tends to zero. For a dirty
sample with pairs created by disorder it is expected that this
intercept remains finite. Thus, we conclude that these are
observations of dissipation due to vortex-antivortex pairs that
are excited by quantum phase fluctuations. These are quenched at
higher fields, thus recovering the LO regime. The mechanism  that
causes dissipation is therefore the dissociation of these pairs.

In order to understand the low current data we first fit the observed
voltage as a power of the current

\begin{equation}
\label{vi} V(I)=V_0[(I/I_c)^{\alpha} -1]
\end{equation}

\noindent The nonlinear regime before the instability is not
hysteretic and fits well a power law above $I_c$ giving $\alpha
\equiv z+1= 2.5$ (while $\alpha =2.5$ gives best fit, a reasonable
fit can be obtained with $\alpha$ in the range 2.4 to 3.0), and
$I_c = J_c  w  d  \simeq 0.008$ A is a critical current associated
with the onset of strong nonlinerarity due to vortex pair
formation  ($w$ is the width of the sample, $d$ its thickness).
Here we introduce the dynamical exponent $z$, which is found to be
smaller than 2. Note that $\alpha$ is equal to 3 ( {\it i.e.}
$z=2$) in classical Kosterlitz-Thouless theory \cite{kt} close to
$T_{KT}$ \cite{nelson}. Using the data presented in figure 1,
figure 4 depicts $V$ $vs.$ $I$. This figure shows that there is an
appreciable deviation from equation \ref{vi} for $I$ near $I_c$
indicating the presence of a population of pairs with spacing
greater than the average spacing. To understand $I_c$ and its
relation to the average spacing between vortex pairs, we note that
the energy of a vortex pair with separation $r$ in the presence of
a current density $J$ is:

\begin{equation}
\label{er} E(r)=U_{v}ln(r/\xi) + 2E_c -(1/c)J d \Phi_0 r
\end{equation}

\noindent where $U_v=4k_BT_{KT}$ \cite{pierson} is the Kosterlitz-Thouless
transition scale \cite{kt} and $E_c=0.39U_v$ \cite{hu} is the core
energy. Minimization gives $r_{av}=cU_v / J_c d \Phi_0 $. Thus, $I_c$
found in figure 4 implies $r_{av} \simeq 105$ {\AA}, or a size
equivalent to a few vortex cores. If we were observing a
thermal-driven Kosterlitz-Thouless transition \cite{minnhagen},
the non-linear $I-V$ behavior before the instability would
originate in thermally-activated hopping over the barrier
$E(r_{av})$. Choosing $\xi=20$ {\AA} in equation \ref{er}, this rate can
be calculated for the temperatures at which the data was taken. We
find that at $55$ K the rate is suppressed by a factor
5x$10^{-4}$, while at $25$ K it is suppressed by a factor
6x$10^{-8}$. Moreover, we note that between $25$ K and $35$ K the
rate would change by a factor of 100, implying a similar change in
the leveling field, while experimentally this change is less than
a factor of two. We therefore conclude that the observed
nonlinear behavior and voltage instability must originate from
vortex-antivortex pairs induced by quantum fluctuations and not
from thermally created pairs. The smaller dynamical exponent can
also point to a quantum mechanical dominated result as all
classical results on high-$T_c$ show increased $z$ above
2 \cite{pierson}.

To model the distribution of pair spacings we assume
a simple harmonic oscillator model for the wave function of the
vortex-antivortex pairs:

\begin{equation}
\label{psi} \psi(r)=re^{-{1 \over 2} (r/a)^2}
\end{equation}

\noindent where $a \simeq r_{av}$ is of order a few vortex core distances.
Then we can relate the distribution of pair spacings to the
distribution of critical currents by writing for fixed $r$

\begin{equation}
\label{vrj} v(r,J)=\rho [(J/J_c(r))^{\alpha}-1]\theta(J-J_c(r))
\end{equation}

\noindent where $\rho$ is the density of pairs.
Using equation \ref{vrj} the voltage response as a function of
$r$ can then be  obtained by integrating over the pair distribution as

\begin{equation}
\label{vj} V(J) \propto \rho\int_{A/J}^\infty dr |\psi(r)|^2 v(r,J)
\end{equation}

\noindent which may be written in dimensionless form as

\begin{equation}
\label{vjj} V(J)=V_0 \int_{A/J}^\infty [ (\tilde{J}x)^{\alpha}-1)] x^2 e^{-x^2}dx
\end{equation}

\noindent Here $\tilde{J} = J/J_c(a)$, $A=(cwU_v/\pi\hbar a)$, and
$eV_0=(1/c)\rho Jd \Phi_0a$ represents the voltage scale
corresponding to dissipation at a length scale $a$. The resulting
$J$ dependence is  parameter free and depends only on the scaling
index $\alpha$. The result for $\alpha=2.5$ is shown in  figure 4
as an inset and fits the data remarkably well. We take this as an
indication that the low current tail in the $I-V$ curves is a
direct measure of the distribution of vortex pairs excited by zero
point motion. Coupled with the high current instability showing
that the resistivity is due to vortex motion, we believe this is
the first direct evidence of quantum excitation of vortex pairs in
high-$T_c$ superconductors at temperatures well below $T_c$. This
lends support to the idea that the loss of phase coherence in
high-$T_c$ superconductors may be directly attributed to
quantum-fluctuation-induced winding of the superconducting order
parameter \cite{doniach1,kivelson}.

Observation of large dissipation in high-$T_c$ films at low
temperatures was previously reported by Corson {\it et al.}
\cite{corson2}, who measured complex conductivities in the
frequency range of 100-600 GHz of BSCCO films from the same source
as those used in this work. At low temperatures they found a
residual real-part conductivity that scales with the $T=0$
superfluid density and is larger than any contribution that could
be due to thermal population of quasiparticles at the d-wave gap
nodes.

To ascertain the relevance of the superconductor pairing symmetry
to our conclusions, we performed similar measurements on $120$
{\AA} film of amorphous MoGe previously used to study vortex
dynamics by Hellerqvist {\it et al.} \cite{hellerqvist}. These
films were chosen for their very short coherence length, large
penetration depth, and relatively weak pinning, similar to very
thin high-$T_c$ films \cite{kapitulnik}. The observed behavior of
the MoGe films resembles that of the BSCCO films; the inset in
figure 3 shows a $V^*$ curve at $T=4.3$ K ($T_c=6.6$ K). Again,
the extrapolation of $V^*$ as a function of $H_{\perp}^{1/2}$
gives zero dissipation at zero field. However, it is clear that
this system is not a d-wave superconductor, thus the dissipation
is wholly due to vortices. Note also that the density of
vortex-antivortex pairs is rather high. The low temperature value
can be extracted from the crossover field. At $T=25$ K this field
is approximately 0.25 T, of order 0.5\% of $H_{c2}$, indicating a
pair density of similar value. The application of a finite field
that creates vortices with only one sign therefore suppresses the
quantum fluctuations as expected.

Finally, we would like to comment on the consequences of the
$H^{1/2}$ dependence of $V^*$. It is interesting to connect our
findings to other measurements related to the size of the vortex
core, a key feature in the LO theory. Note that LO's result is
derived from the calculation of the velocity at which the
viscous-friction force has a maximum. Their expression for the
critical velocity has the form $v^{*2}_{\varphi} \sim D/\tau _{in}
\sim {\xi ^2 /(\tau _{in}\tau _{GL})}$, where $\tau _{GL}$ is the
fundamental Ginzburg-landau time. As noted above, at low fields $
v^*_\varphi \sim {(\Phi _0 / B)^{1/2}/ \tau_{in}  }$. Thus we may
conclude that in this regime $\xi \sim l_H \sim H^{-1/2}$ and that
this apparent increase in vortex core size is a consequence of the
fact that the quasiparticle distribution shows a strong spatial
inhomogeneity with portions of the superconducting phase still
displaying the equilibrium distribution. Similar behavior, an
apparent increase in vortex core size, was also observed in recent
muon-spin-rotation spectroscopy experiments on high-$T_c$
\cite{sonier1} and conventional layered superconductors
\cite{sonier2}, as well as in specific heat measurements
\cite{sonier3}.  We believe that all these measurements are in
fact observing the strongly inhomogeneous quasiparticle
distribution in the system at low fields.

In summary, we present evidence that in thin films of BSCCO, at
low temperatures and low magnetic field (including $H=0$ T), there
exists anomalous dissipation that is similar to vortex dissipation
observed at finite magnetic fields. This provides direct evidence
for the existence of vortex-antivortex excitations due to quantum
fluctuations of the phase of the superconductor.

We thank Susanne Doettinger and John Clem for many useful
discussions. Work supported by AFOSR grant AF49620-98-1-0017.
JC thanks DoD fellowship support.

\figure{FIG. 1. Typical I-V measurement cycle. The current is ramped
in a triangular fashion and the corresponding voltage is
recorded. $V^*$ marks the onset of the instability (see text).
\label{fig1}}

\figure{FIG. 2. Instability voltage as a function of angle for
several magnetic fields at $T=35$ K. Inset shows data collapse
when plotted against the perpendicular (to the sample) component
of the magnetic field. \label{fig3}}

\figure{FIG. 3. Instability voltage vs. $H^{1/2}$ at two different
temperatures ($T_c=82$ K). Inset shows similar data for a $120$
{\AA} sample of MoGe (see text). \label{fig4}}

\figure{FIG. 4. The data of figure 1 plotted as $V$ vs. $I$. Insert
shows the fit to equation \ref{vjj}. $I_c$ is discussed in the text.
\label{fig2}}

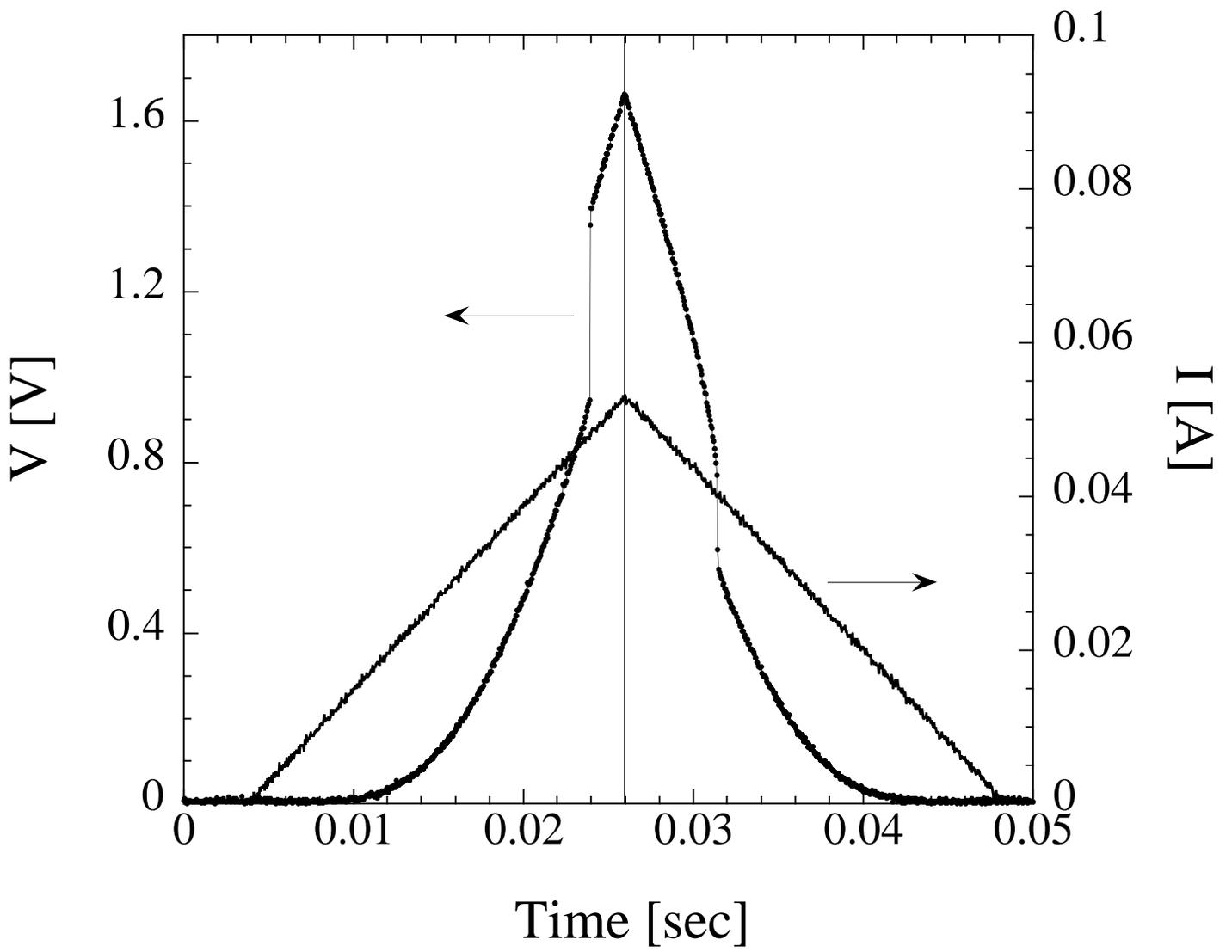
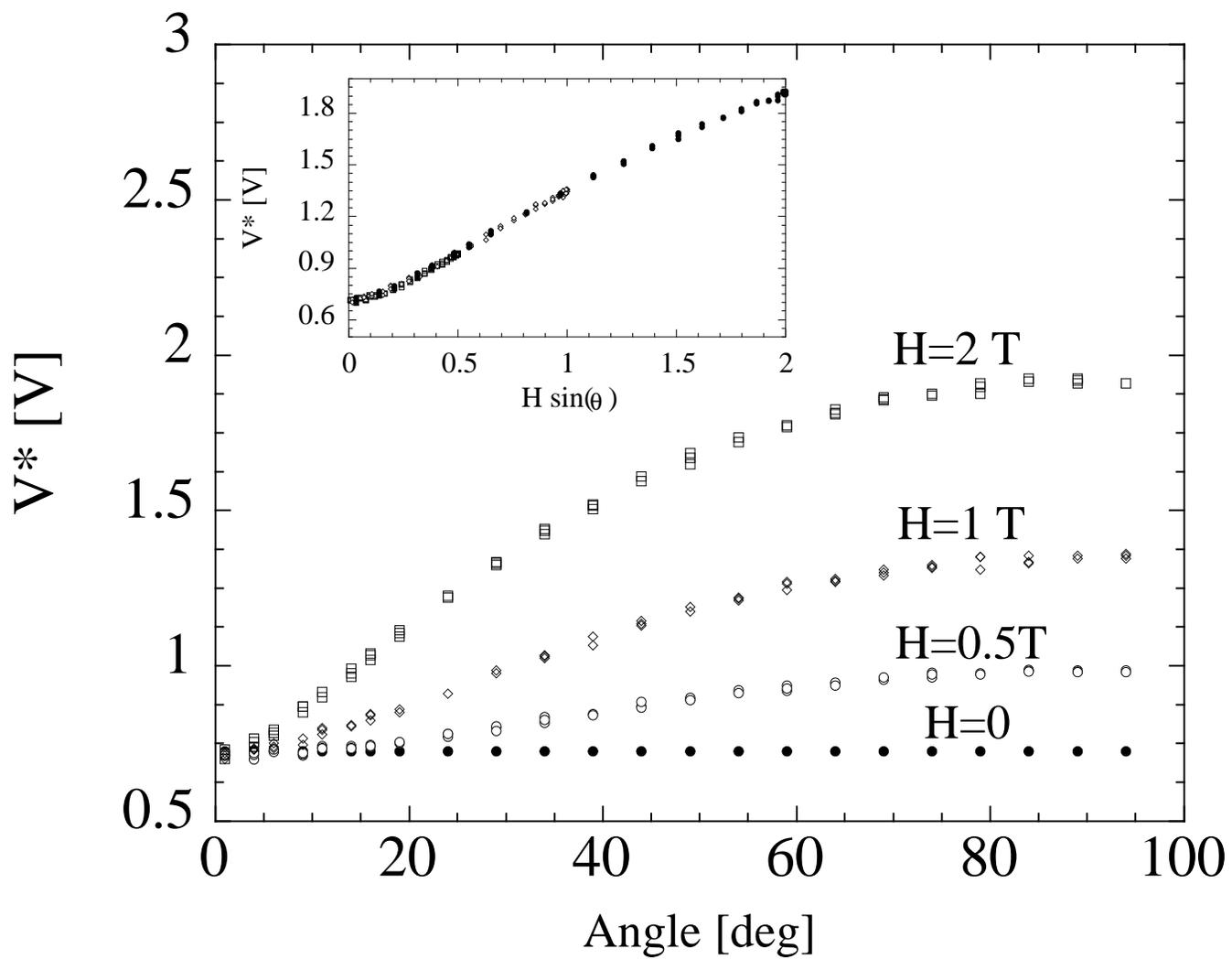
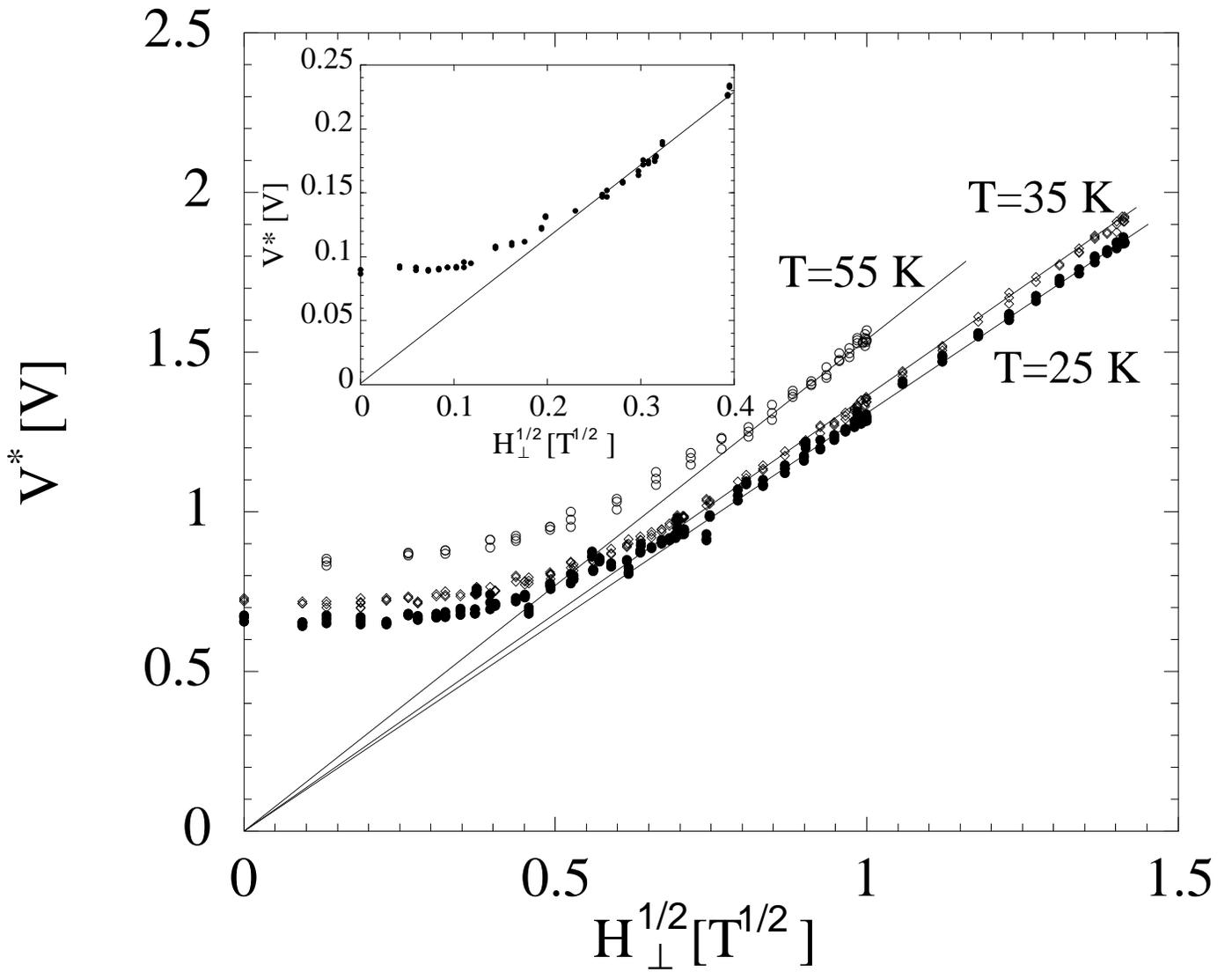
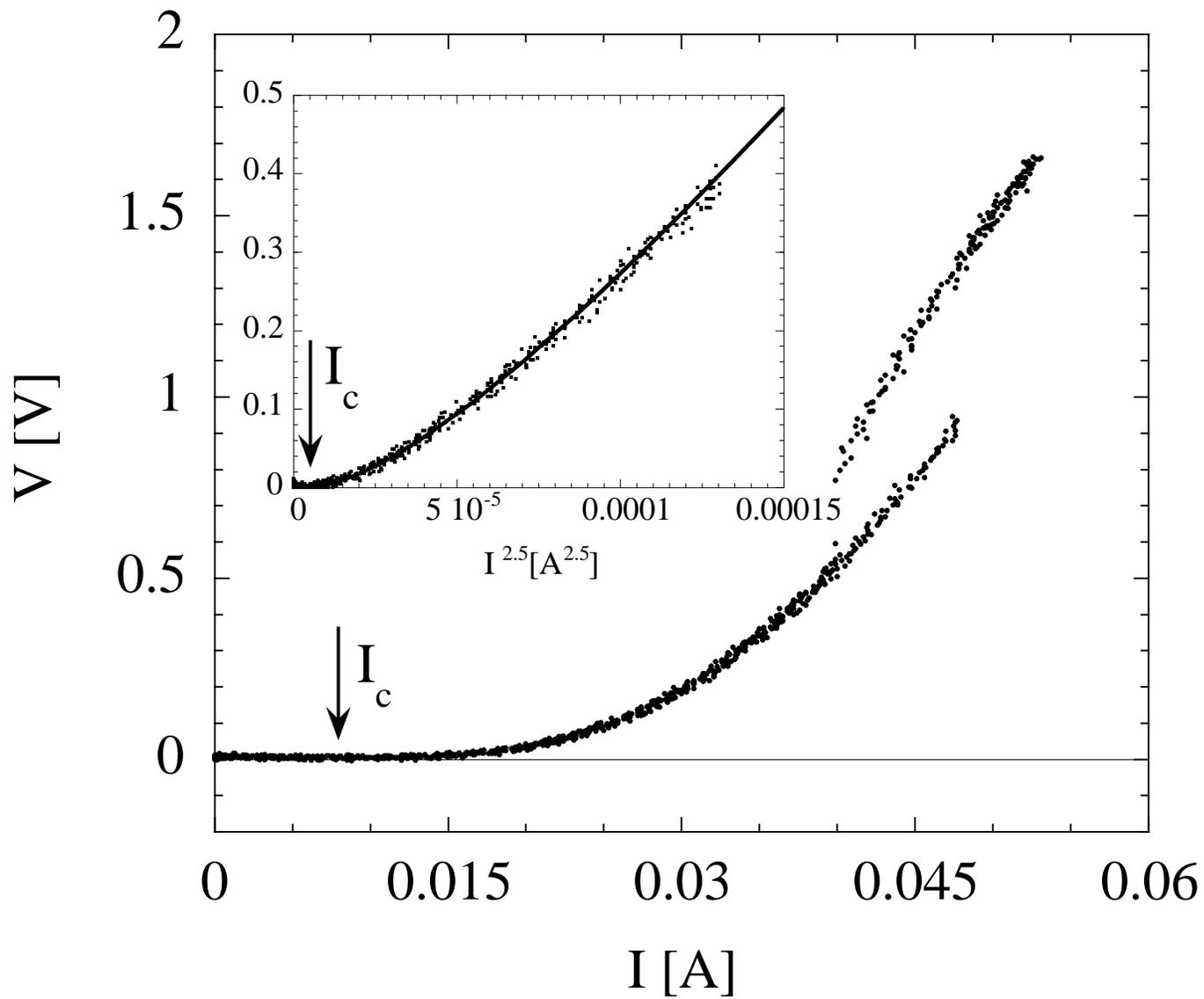
\end{document}